\input psfig
\documentstyle[psfig]{article}
\begin{document}
%
{\hfil\hfill \bf EFI 95-78}
%
%
\vspace{3mm}
\begin{center}
{\bf DETECTION OF ATMOSPHERIC CHERENKOV RADIATION}
\vspace{2mm}
{\bf USING SOLAR HELIOSTAT MIRRORS}
\vspace{6mm}

R.A. Ong$^1$, D. Bhattacharya$^2$, C.E. Covault$^1$,

D.D. Dixon$^2$, D.T. Gregorich$^3$, D.S. Hanna$^4$, S. Oser$^1$,

J. Qu\'ebert$^5$, D.A. Smith$^5$, O.T. T\"umer$^2$, and A.D. Zych$^2$ 

\vspace{4mm}

{\it $^1$The Enrico Fermi Institute,
The University of Chicago, 

Chicago, IL 60637, USA

\vspace{2mm}

$^2$Institute of Geophysics and Planetary Physics,
The University of 

California, Riverside, CA 92521, USA

\vspace{2mm}

$^3$IPAC, California Institute of Technology, Pasadena, CA 91125 and

Dept. of Physics and Astronomy, California State 
University LA, 

Los Angeles, CA 90032, USA

\vspace{2mm}

$^4$Department of Physics, McGill University

Montreal, Quebec H3A 2T8, CANADA

\vspace{2mm}

$^5$C.E.N. de Bordeaux-Gradignan, Le Haut Vigneau,

Gradignan 33175, FRANCE}

\vspace{6mm}

(To be published in Astroparticle Physics)

\vspace{7mm}

\end{center}


\section*{Abstract}
There is considerable interest world-wide in developing
large area atmospheric Cherenkov detectors for ground-based
gamma-ray astronomy.
This interest stems, in large part, from the fact that the
gamma-ray energy region between 20 and 250 GeV is 
unexplored by any experiment.
Atmospheric Cherenkov detectors offer a possible
way to explore this region, but large photon
collection areas are needed to achieve low energy thresholds.
We are developing an experiment using the heliostat mirrors of
a solar power plant as the primary collecting element.
As part of this development, we built a detector using
four heliostat mirrors,
a secondary Fresnel lens, and
a fast photon detection system.
In November 1994, we used this detector to
record atmospheric Cherenkov radiation produced by
cosmic ray particles showering in the atmosphere.
The detected rate of cosmic ray events was consistent with an
energy threshold near 1 TeV.
The data presented here represent the first detection of atmospheric
Cherenkov radiation using solar 
heliostats viewed from a central tower.

\section{Introduction}

The instruments on the Compton Gamma Ray Observatory (CGRO) have
revealed that the gamma-ray sky is rich and exciting.
Over 100 point sources of gamma-rays have been detected
by the EGRET detector on the CGRO at energies up to 20 GeV 
\cite{ref:CGRO}.
Gamma-rays are detected from objects such as bursts,
pulsars (e.g. Crab and Vela), 
active galactic nuclei (e.g. 3C279, Mrk 421), and nearby galaxies
(e.g. LMC), but the majority of the sources are
still unidentified with known objects.
Largely due to its limited collection area, the EGRET instrument
lacks sufficient sensitivity to detect sources
above 20 GeV.

Ground-based telescopes 
have detected TeV gamma-rays from a handful of objects
(e.g. Crab, Mrk 421).
The detections clearly demonstrate that gamma-ray astronomy can
be done from the ground using the atmospheric Cherenkov technique 
\cite{ref:Snowmass}.
However, the current generation of atmospheric Cherenkov telescopes
is limited by signal-to-noise considerations to operate at energies
above 250 GeV.
Therefore, an important goal in gamma-ray astronomy
is the exploration of the 
energy region between 20 and
250 GeV.
This region is one of the last remaining windows of
the electromagnetic
spectrum where no observations have been made.
We would like to search this window for possible new astrophysical
phenomena, and would 
like to connect observations made at low energies
by satellite experiments (such as EGRET) with ones made at high energies
by ground-based observatories.
These are the most important 
scientific motivations for the 
development of gamma-ray telescopes in this energy regime 
\cite{ref:NRC}.

\subsection{Atmospheric Cherenkov Technique}

Although the Earth's atmosphere is opaque to gamma-radiation,
the products of extensive air showers created by 
gamma-rays can be detected at ground level.
High energy gamma-rays interact in the upper atmosphere
and produce electromagnetic cascades consisting largely of
electrons, positrons, and photons.
The relativistic electrons and positrons emit Cherenkov radiation
which is beamed to the ground.
The Cherenkov photons hit the ground with a narrow spread in arrival
times ($< 40\,$nsec) and form a circular light pool approximately
250$\,$m across.
Atmospheric Cherenkov telescopes, using mirrors and fast photomultiplier
tubes (PMTs), detect these photons to reconstruct the energy and direction
of the primary gamma-ray.
The atmospheric Cherenkov technique has been successfully used
for gamma-ray astronomy at energies from 250 GeV to 10 TeV by several
observatories, such as Whipple (U.S.), Cangaroo (Australia), and
Th\'emis (France).

Although there are a large number of Cherenkov photons reaching the
ground in an air shower,
the photons are spread out over a large area, leading to
small densities at any given ground location.
For example, a shower created by a vertically incident 50 GeV gamma-ray
contains, on average,
more than 200,000 photons at ground level.
However, the mean density
within 150$\,$m of the shower axis is
only $\sim 2.6$ photons/m$^2$.
Small density values require large mirror collection areas in order
to concentrate enough Cherenkov light 
onto the detecting elements of a telescope.

The energy threshold of a Cherenkov telescope is set by its
ability to trigger on a signal of Cherenkov photons amidst
the background of night sky photons.
If we assume that the energy threshold, $E_{\rm{th}}$, 
depends on the minimum signal-to-noise
ratio required for triggering, we can derive the approximate
way in which the threshold depends on the telescope parameters:

\begin{equation} 
\ E_{\rm{th}} \ \propto\      
  \sqrt{\,B\,\Omega\,t\,/\,A\,\epsilon\,}\ .  \label{eq:threshold}
\end{equation}

\noindent Here $B$ is the flux of night sky photons,
$\Omega$ is the solid angle viewed by the detecting element,
$t$ is the trigger time window.
$A$ is the telescope mirror area, and $\epsilon$ is the
efficiency of light collection 
\cite{ref:Weekes}.
We can reduce the
energy threshold of a Cherenkov telescope 
by minimizing the background light level, the field-of-view, and the
trigger time window.
or by maximizing the photon collection efficiency and the
mirror area.
There are practical limitations on reducing the field-of-view
or time window that are imposed by the physics of the shower.
A minimimum field-of-view of $0.5^\circ$ (full angle) is required
to contain most of the Cherenkov image.
The time window cannot be reduced much below a few nsec, which is
the characteristic width of the Cherenkov pulse.
The most straight-forward way to lower the energy threshold
is to increase the amount of mirror collection area.

Ground-based telescopes using the air shower technique
must contend with a large background of showers initiated
by isotropic cosmic rays.  This background flux is typically
two to three orders of magnitude larger than gamma-ray fluxes.
In order to reliably detect high energy gamma-rays, 
Cherenkov telescopes must reject a significant fraction of the
cosmic ray showers.
At primary energies below 100 GeV, substantial rejection
comes from the fact that cosmic ray showers contain
much less Cherenkov light than showers initiated by
gamma-rays. In cosmic ray showers, a large fraction of the total
energy goes into products (hadrons, muons, neutrinos) which do
not produce Cherenkov radiation, whereas in gamma-ray showers, 
all the energy goes into the electromagnetic cascade.  

A second observation is that
showers produced by cosmic rays are much more chaotic and
irregular in development than those initiated by gamma-rays. 
These development characteristics translate
into significant differences between gamma-ray and cosmic ray showers
in the angular and lateral 
distributions of the Cherenkov light.
Atmospheric Cherenkov detectors that exploit these shower features
are able to reject a substantial fraction of the cosmic ray showers.
For example, the Whipple Observatory has used an imaging technique to
reject over 99\% of the cosmic ray background.
This technique led to the first convincing ground-based
detection of gramma-rays \cite{ref:Vacanti}.

\subsection{Large Area Cherenkov Detectors}

Current state-of-the-art atmospheric Cherenkov telescopes
have mirror areas $\sim 100\,$m$^2$, and energy thresholds
of $\sim 250$ GeV.
These experiments sample only a small fraction of the available
Cherenkov light.
Recently, designs for a number of 
new large area Cherenkov detectors have been suggested.
One design concept (VHEGRA) makes use of large
(10\,m diameter) reflectors in which 
each reflector has its own PMT
camera \cite{ref:Akerlof}.
Another group (Telescope Array) is developing a telescope
consisting of more than one hundred small ($3\,$m diameter)
reflectors \cite{ref:Teshima}.
In this design, each reflector would have a camera of sixteen multi-anode
PMTs.
A third design concept consists of one (or more) large 
hemispherical reflectors or ``bowls'' constructed
by filling the insides of a cavity in the ground with mirrors 
\cite{ref:Ahlen}.
In the most ambitious version of this design,
the cavity would be 300 to 400$\,$m across, and
a large camera, suspended at the mirror focal point, would
follow astronomical sources by rotating to view different parts
of the sky.

In each of the design concepts mentioned above, the
price of the reflectors themselves dominates the overall cost
of the experiment.
We are developing an atmospheric Cherenkov detector that makes
use of the large mirror area available at central tower solar plants.
The great advantage here is that the mirrors have already been built
and are available.

\subsection{Solar Array Cherenkov Detectors}

The idea to use large solar arrays for collecting Cherenkov radiation
from air showers was introduced more than a decade ago with application
to the solar reflector (heliostat) field
of the National Solar Tower Test Facility (NSTTF)
at Sandia National Laboratories (Albuquerque NM, USA) 
\cite{ref:Danaher}.
The original design concept was to reflect Cherenkov light from 
heliostats onto a camera of PMTs located on top of the
central tower, viewing the heliostat field.
This design had a number of technical problems:
1) the large 
($>1\,$m diameter) Cherenkov image sizes 
produced by the
heliostats required very large PMTs,
2) each PMT saw light from more than one heliostat, and
3) designing a detector trigger was difficult because the
time of propagation for light from a heliostat to
the tower varied, depending on the position of the heliostat
and on the inclination angle of the shower.

In 1990, members of our group proposed a gamma-ray observatory
using the large heliostat
array of the Solar Two Power Plant (Daggett CA, USA) 
\cite{ref:Name,ref:Tumer}.
The proposal suggested the use
of an optical secondary located on the
central tower.
Here, the secondary (either a large mirror or Fresnel lens) serves two
important tasks.
First, it acts to focus the light from each heliostat down to a small
spot that can be collected by a conventional (1-2{\tt "}) PMT.
Second, it creates an image of the heliostat field in the focal plane of
the secondary. 
Each heliostat is imaged to an unique location in the image plane.
In this way, each PMT sees the light from only one heliostat.
The use of secondary optics therefore solves the first two problems
associated with the original design concept \cite{ref:Danaher}.
The problem
of varying time delays remains, but the use of secondary optics
allows each PMT to have its own unique delay circuit.

To form an overall experiment trigger, the PMT signals from
the heliostats would be selectively delayed and combined.
Two possible delay schemes can be imagined.
The first scheme would delay the PMT
analog signals by means of good quality coaxial cables.  The
variation in delay lengths would be achieved by programmable
switches, and a trigger would be formed from the analog sum of the signals.
The second scheme would discriminate the PMT signals
at low levels.
The digitally delayed
discriminator outputs would be combined to form an overall trigger.

In addition to lowering the energy threshold, the use of an array of
mirrors also
permits measurement of the lateral distribution of the Cherenkov light.
As discussed earlier, this distribution can be used to provide separation
between showers initiated by gamma-ray and cosmic ray (hadronic)
primaries.
We are currently examining the degree of separation
that could be achieved from a hypothetical large area
Cherenkov array using Monte Carlo simulations.

Since 1993, considerable interest has arisen in the possibility of
atmospheric
Cherenkov telescopes built around solar heliostat arrays.
There are now two groups developing prototype experiments based
on this design concept.
Some of us (JQ and DAS) are involved with a French group
initiating an experiment to use up to 180
heliostats of the Th\'emis solar power plant near Targasonne
in the eastern Pyr\'en\'ees \cite{ref:Pare}.
The remainder are
developing a prototype experiment to use either the Solar
Two Power Plant or the National Solar Tower Test Facility.
To understand the feasibility of the design concept, we have carried
out a program of tests at the Solar Two site.
The results from this program, which started in April 1994, are the subject
of this paper.

\subsection{Solar Two Power Plant}

The Solar Two Power Plant
was constructed from 1979 to 1992, and is
located $20\,$km east of Barstow CA, USA
(34.9$\,$N, 116.3$\,$W, 593$\,$m above sea level).
Solar Two consists of 1818 heliostats spread out around
a central tower, covering an area of approximately 200,000$\,$m$^2$.
Each heliostat has its own altitude-azimuth drive and 
local circuitry for tracking control.
A heliostat mirror is composed of twelve facets with a total reflective
surface of 40$\,$m$^2$.
The heliostats can be controlled from computers at a central location, or
via a specialized interface using a lap-top 
personal computer.

After a test and evaluation phase, the plant was used for power production
until 1988. 
Recently, a consortium of private and government sponsors formed to
refurbish the plant and to resume solar energy production.
The refurbishment of the plant started at the end of 1994,
and was completed in November 1995.
Among other improvements carried out, 
each heliostat was inspected, (if necessary) repaired, and
brought into good optical alignment.
The refurbishment of the plant has therefore had an important beneficial
impact on its potential as an astronomical site.
The Solar Two facility will resume operations for solar power in early
1996 as part of a several-year program.

\section{Site Testing Program}

We now describe the technical progress that has been made on our design
concept.
In April 1994, we started a testing program at the Solar Two site.
The initial goals of this program were to:
1) characterize the properties of the heliostats,
2) determine the suitability of the site for astronomical purposes,
and
3) demonstrate that Cherenkov radiation can be reliably detected 
above background using solar heliostats.
We accomplished each of these initial goals during tests carried out in
1994.

\subsection{Heliostat and Site Characterization}

In April and August 1994, we spent ten days at the Solar Two site
studying the performance of the heliostats and making measurements of
the night sky brightness.
The results from these studies have been presented elsewhere 
\cite{ref:Ong},
and therefore, are only briefly discussed here:

\begin{enumerate}
\item The tracking and pointing accuracies of the
heliostats were measured directly using a laser.  The heliostats track
smoothly to an accuracy of better than $0.05^\circ$.
\item The optical properties of several heliostats were determined
by reflecting sunlight onto a target at the top of the central
tower.  
After suitable alignment of the individual mirror facets,
the heliostat spot sizes at the tower were measured to be
less than 3.0 meters in diameter.  
Since the angular size of the Sun ($\sim 0.5^\circ$) is comparable
to the angular extent of the Cherenkov light in a shower,
the measured spot sizes
determine the optimal diameter of the secondary.
\item Direct photometric measurements of the night sky
brightness due to local light pollution
were made from the triggering rate of 
a collimated PMT.
Although light pollution near the horizon
was high, the brightness looking down into the heliostat field
(where the secondary optics would point) was within 50\% of that found
at a dark location such as Dugway, Utah, site of the
Chicago Air Shower Array (CASA) and Fly's Eye experiments.
\end{enumerate}

The results from these studies encouraged us to build
a detector to see atmospheric Cherenkov radiation.

\section{Atmospheric Cherenkov Radiation Detection}

\subsection{Detector}
\begin{figure}
\centerline{\ \psfig{figure=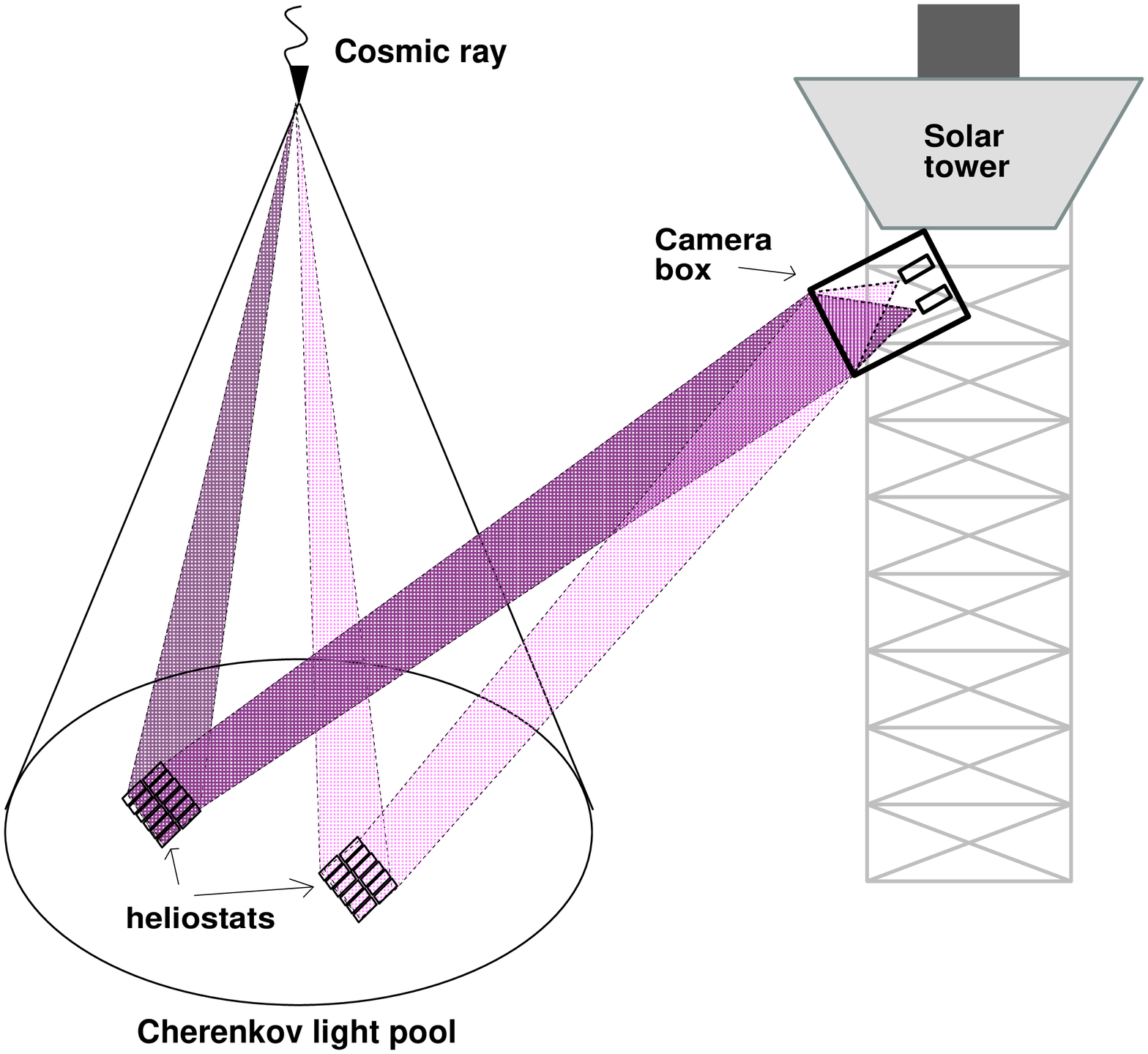,height=13.0cm}}
\caption{Schematic representation of the detector.
A high energy cosmic ray creates an extensive air shower
in the atmosphere. 
The Cherenkov radiation in the air shower is beamed to the
ground where it is reflected by heliostats to
a camera on the central tower.
The camera (located $70\,$m above ground level)
consists of a large Fresnel lens and an
array of photomultiplier tubes.
The camera box is not drawn to scale.} \label{fig:schematic}
\end{figure}

A schematic representation of the detector is shown in
Fig.~\ref{fig:schematic}.
High energy cosmic rays interact in the upper atmosphere
($\sim 10\,$km altitude) to create extensive air showers.
The Cherenkov light
in the shower propagates to ground level.
The primary optic consisted of four solar heliostats
with a total reflective area of $160\,$m$^2$.
The heliostats reflected Cherenkov radiation to a
camera box on the central tower.
The camera consisted of a secondary optic (Fresnel lens)
and photomultiplier tubes (PMTs).
The PMT signals were electronically recorded by a
data acquisition (DAQ) system.

\subsubsection{Heliostat Selection}

\begin{figure} 
\centerline{\ \psfig{figure=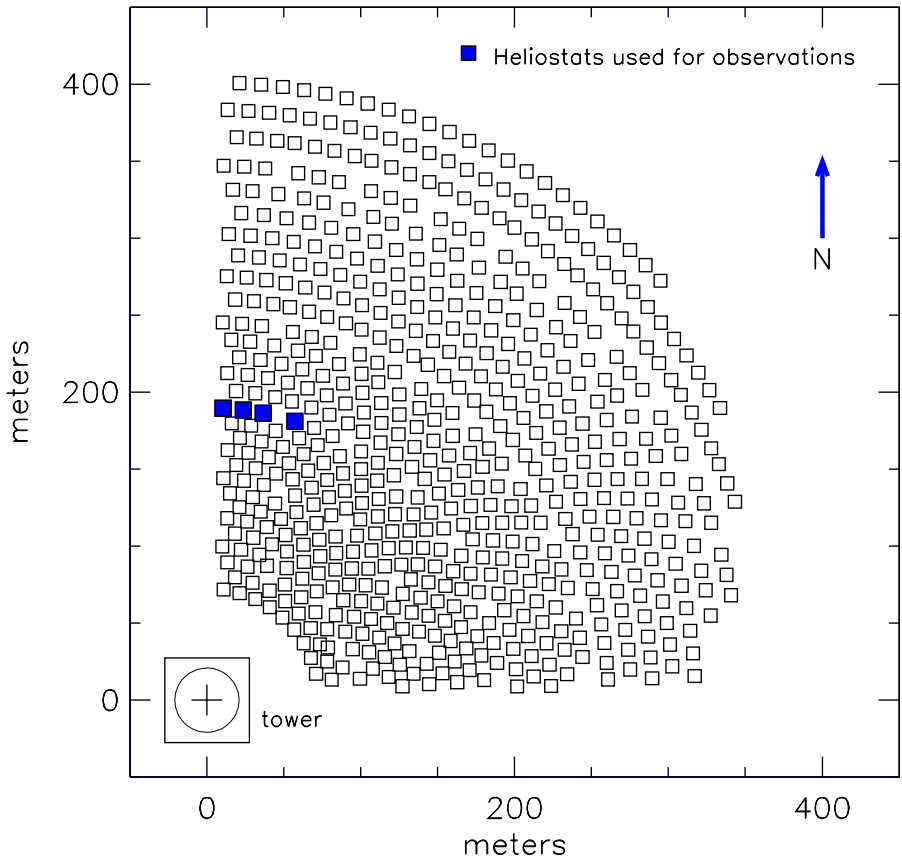,height=14.0cm}}
\caption{
Northeast quadrant of the Solar Two heliostat field.
Each square represents a heliostat consisting
of a $40\,$m$^2$ mirror
and an altitude-azimuth drive.
The four heliostats used in this work
are indicated by the shaded squares.} \label{fig:field}
\end{figure}

We selected four heliostats \cite{ref:Heliostats}
in the northeast quadrant of the
Solar Two heliostat field, as shown in Fig.~\ref{fig:field}.
The heliostats were approximately halfway out
into the field and were chosen 
to lie along a circular arc so that 
each heliostat was at the same approximate distance from the tower.
In this way, 
there were no relative differences in the light propagation times from
the heliostats to the tower
for vertically incident showers.
The mirror facets of the selected heliostats were washed and
aligned using the image of the Sun on a large target on the tower.
The azimuth and elevation headings of the heliostats were controlled using
a custom-made interface module \cite{ref:Stimulator}
and a lap-top personal computer.

\subsubsection{Camera}

The camera box was made from an aluminum frame of dimensions
{32{\tt "} x 42{\tt "} x 84{\tt "}}.
Aluminum panels painted black covered the frame to make a
light-tight enclosure.
A shutter attachment was mounted at one end
of the box (along the major axis).
This attachment held a {\tt 31" x 41" x 3/16"} Fresnel lens
(Edmund Scientific G31,139) and a removable wooden shutter.
The lens had a focal length of {\tt 40"} and a typical transmission of 85\%
for wavelengths between 400 and 1000 nm.
Four fast {\tt 2"} photomultiplier tubes (Hamamatsu R2154-UV)
were mounted at the focal plane of the Fresnel lens so that the
images of four heliostats were focused onto the PMTs
(one heliostat per PMT).
The PMTs were operated with the cathode at negative high voltage
(typically -1.05 kV), and with a typical current amplification of
$\sim 5 \times 10^5$.
The high voltage was provided by a 32-channel programmable supply 
(LeCroy 4032).

\subsubsection{Electronics}

\begin{figure} 
\centerline{\ \psfig{figure=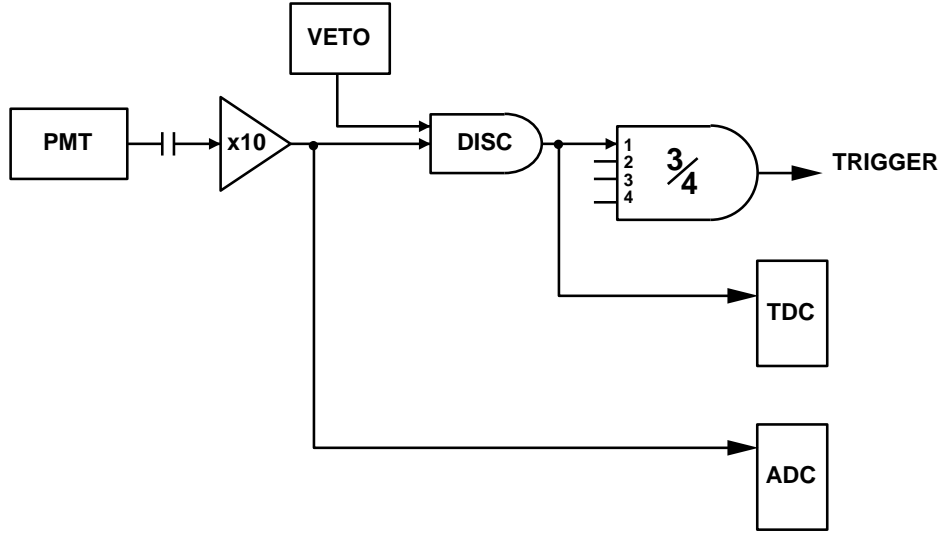,height=7.0cm}}
\caption{
Functional diagram of the electronics used
(one of four channels).
The photomultiplier tube (PMT) signal is amplified (x10)
and discriminated.
The discriminated signal is routed into the trigger logic.
A trigger is formed when any three out of four PMTs
are above threshold within a window of $40\,$ nsec.
The PMT time-of-arrival is recorded in a time-to-digital
converter (TDC), and its pulse-height is recorded in
an analog-to-digital converter (ADC).
A veto circuit is used to eliminate pulses of long duration
created by light from artificial sources.} \label{fig:electronics}
\end{figure}

A schematic diagram of the electronic circuitry is shown
in Fig.~\ref{fig:electronics}.
The electronics consisted of commercial modules using the NIM and CAMAC 
standards.
The PMT signals were carried to the electronics via $15\,$m of
coaxial cable (RG-58A/U).
The signals were AC-coupled through amplifiers (LeCroy 612AM)
with a signal gain of ten.
We used amplifiers in order to keep the PMT gains below
$10^6$.
The amplified signals were delayed by 
$70\,$m of coaxial cable (RG-8/U), and their charges
were measured by 
analog-to-digital converters (ADCs).
The ADCs (LeCroy 2249W) had 11-bit resolution, a conversion
scale of $0.25\,$pC/count, and were gated by a $40\,$nsec wide pulse.

The amplified signals were also discriminated (LeCroy
623B) at discriminator levels between $90\,$mV and $225\,$mV.
The discriminated outputs were used to form a trigger coincidence,
and to stop time-to-digital converters (TDCs).
The TDCs (LeCroy 2228A) had a dynamic range of 11-bits with a least
count value of $250\,$psec.
The trigger logic (LeCroy 365) required the overlap of at least
three (out of four) $40\,$ns wide pulses.
The trigger rates for three and four-fold coincidences, as well
as the PMT single counting rates, were recorded by
scalers (LeCroy 2551).

An important part of the electronic circuitry was a veto circuit
designed to eliminate the effects of background light
produced by man-made sources.
The most significant sources of this type were aircraft warning
beacons at a neighboring power plant (located 3\,km
southwest of the Solar Two site).
These beacons produced flashes of light which were intense, but which occurred
at a low repetition rate ($\sim$ 1 Hz).
The PMT pulses caused by the flashes
were much longer in duration than Cherenkov signals ($100\,\mu$sec
versus $10\,$nsec).
Therefore,
a veto circuit based on pulse length could be used to inhibit the trigger
logic during noise pulses.
The veto circuit was made by splitting the amplified signal from each
PMT into two paths. 
Along one path, the signal was delayed through $30\,$m of coaxial
cable (RG-58A/U).
Along the other path, an updating discriminator
(Phillips 711) was used to indicate the presence of a signal
that remained over threshold for more than 100\,nsec.
In this manner, all spurious long pulses 
were eliminated from the trigger.

\subsubsection{Data Acquisition}

The trigger signal generated a gate for the ADCs and a common start for the
TDCs.
The data recorded in the ADCs and TDCs were digitized and read out
through a 16-bit parallel CAMAC interface (Kinetic Systems 3922) to
an Intel 286-based personal computer.
The data were acquired on a local hard disk and then transferred
via modem to a remote Unix workstation.

\subsection{Operating Conditions}

The data used in this work were taken on two moonless
nights (November 4 and 5, 1994).
On the first night, the weather conditions were good, with clear
skies and a wind speed less than 20 mph at ground level.
On the second night, there was a significant amount of high level haze.
Before the tests started, we studied the pointing
of each heliostat by locating
the image of the Sun on the tower.
We recalibrated the absolute 
heliostat pointing for each night of operation.

At the start of nighttime operation, high voltage was applied
to the PMTs, but the camera shutter was kept closed.
The PMT single counting rates and three-fold and four-fold coincidence
rates were measured for a range of
different discriminator settings.
We repeated these measurements with the camera shutter open,
but with the heliostats oriented randomly in the sky
(i.e. with the heliostats not pointed to a common location).
Finally, the heliostats were aligned to a common point in the sky.
The trigger rates and pulse-height distributions were recorded for
various discriminator settings and for various heliostat zenith angle
configurations.

\subsection{Results}

\subsubsection{Trigger Rates}

\begin{figure} 
\centerline{\ \psfig{figure=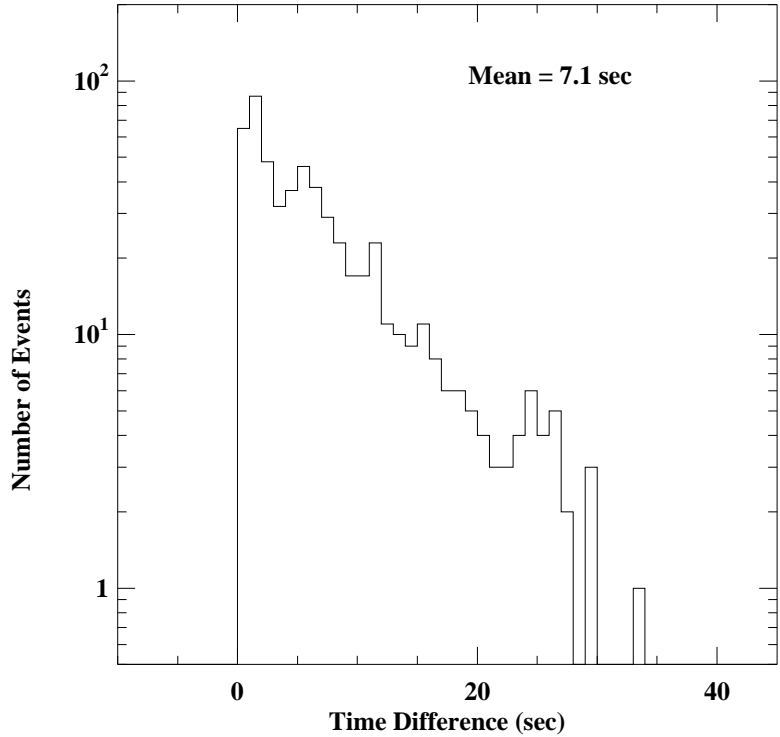,height=12.0cm}}
\caption{
Distribution of time intervals between successive events.
Data shown were taken with the detector triggering on vertical 
air showers
on November 4, 1994, at a PMT threshold of 125\,mV.
The data are well described by a single exponential function with
mean ($7.1\pm 0.5$)\,sec.
The mean agrees with the observed event 
rate of 9.1/minute.} \label{fig:times}
\end{figure}

The PMT single counting rates were estimated from scaler counters that
monitored each PMT.
The three and four-fold coincidence rates were calculated from
additional scaler counters or from the TDC information.
The two methods used to estimate the coincidence rates gave identical
results.
In addition,
the distributions of the time intervals between successive
events predicted coincidence rates that were consistent with the
measured rates.
For example, Fig.~\ref{fig:times} shows the time interval distribution of successive
events for data taken on November 4.
The shape of the distribution for small time differences indicates
that the detector deadtime was small ($< 5\%$).

Table~\ref{tab:pmtrates} 
shows typical PMT single counting rates and 
three and four-fold coincidence 
rates for the detector under various operating conditions.
The coincidence rates demonstrate that the detector was
triggering on atmospheric Cherenkov radiation since
coincidences only occurred when the heliostats were aligned
to a common point in the sky.
The fact that 
the coincidence window 
was short ($40\,$nsec) excluded the possibility that artificial
light sources caused accidental coincidences.
The pulses of individual PMT signals in coincidence events
were examined on 
a storage oscilloscope, and were found to have very
sharp rise-times ($\sim 3\,$nsec) and profiles 
consistent with Cherenkov radiation
produced in the atmosphere.

\begin{table}
\caption{Typical
PMT single counting rates and trigger
coincidence values (in 10 minute intervals) for various
detector configurations at a PMT threshold of 125\,mV.
The detector configurations were:
A) high voltage on and shutter closed,
B) high voltage on, shutter open, and heliostats not aligned,
C) high voltage on, shutter open, and three heliostats aligned,
D) high voltage on, shutter open, and four heliostats aligned.
The range in the single counting rates reflects gain differences
among the PMTs.
The difference in the single counting rates between configurations
B and C reflects the fact that the heliostats were pointed to
different sky locations in these two configurations.
For configurations C and D, the heliostats were aligned to detect
vertical air showers.
The three-fold coincidence values include those events which are
also four-fold. }
\label{tab:pmtrates}
\vspace{10pt}
\begin{center}
\begin{tabular}{c|c|c|r}
\hline
Configuration & PMT single count. rate  & 3-fold coin.
& 4-fold coin. \\
\hline
A & 3.6-5.5 Hz\ & None\ \ \ \ \  & None\ \ \ \ \ \\
B & 4.5-6.8 kHz & None\ \ \ \ \ & None\ \ \ \ \ \\
C & 4.8-7.2 kHz & 75\ \ \ \ \ \ & None\ \ \ \ \ \\
D & 4.8-7.2 kHz & 91\ \ \ \ \ \ &  48\ \ \ \ \ \ \ \\
\hline 
\end{tabular}
\end{center}
\vspace{15pt}
\end{table}

Table~\ref{tab:pmtrates} also shows that when a single heliostat
was pointed away from the common point in the sky, the four-fold
coincidence rate went to zero.
This demonstrates that
each PMT saw light from
only one heliostat
and that there was no significant cross-talk from one heliostat
to another.

\begin{figure} 
\centerline{\ \psfig{figure=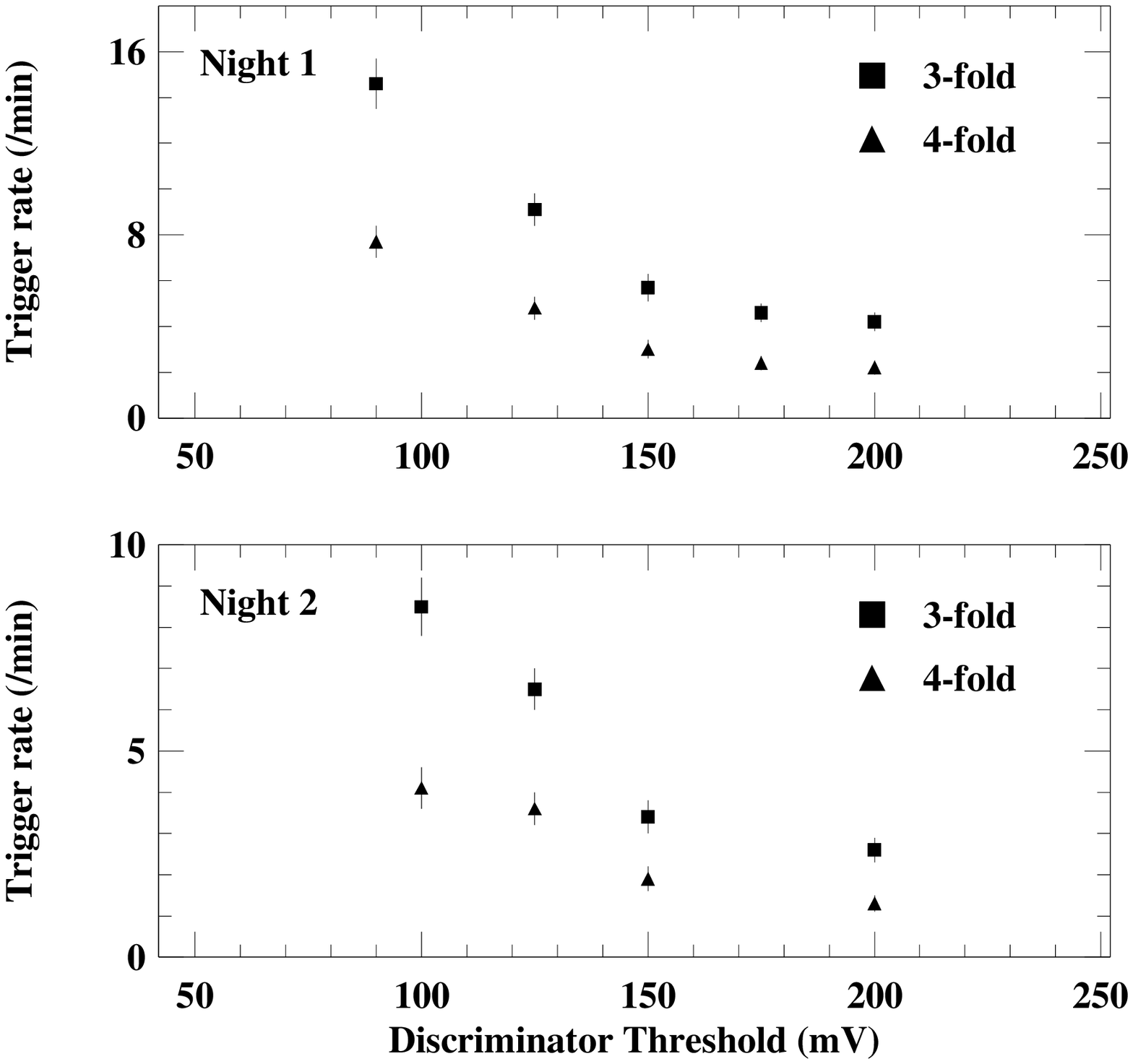,height=14.0cm}}
\caption{
Detector three and four-fold coincidence rates as a function of PMT
discriminator threshold.
The data for observations made on two different nights
are shown for the detector triggering on vertical air showers
(Night 1: November 4, 1994, Night 2: November 5, 1994).
The atmospheric visibility on the second night was degraded 
by the presence of a layer of high haze.} \label{fig:rates}
\end{figure}

Fig.~\ref{fig:rates} shows the three and four-fold coincidence rates
as a function of the PMT discriminator threshold for
data taken on two different nights 
(Night 1: November 4, Night 2: November 5, 1994).
The rates decrease monotonically with increasing discriminator
level, showing that the detector energy threshold rose as the
discriminator level was increased.
The three-fold coincidence rate was a factor of two
larger than the four-fold rate, indicating a significantly
lower energy threshold for the three-fold events.
The coincidence rates observed on the first night were
higher than those observed on the second night for the same
operating conditions.
This change was presumably caused by a reduction in the clarity of
the atmosphere during the latter observations.

Assuming that the detector was triggering on atmospheric Cherenkov radiation
produced by cosmic ray interactions in the atmosphere, we expect
an approximate reduction in the trigger rate of 3.2 for each factor
of two increase in threshold (E$^{-1.66}$ integral spectrum).
For the observations made on the first night, we 
measured reduction factors of 3.0 and 2.9 for the three and four-fold
coincidence data, respectively.
On the second night, we measured factors of 3.3 and 3.2, respectively.
The measured reduction factors 
(with a typical uncertainty of 10\%)
are in good agreement with expectations, supporting the assumption that
the detector was
triggering on 
air showers induced by cosmic rays.

\begin{figure}
\centerline{\ \psfig{figure=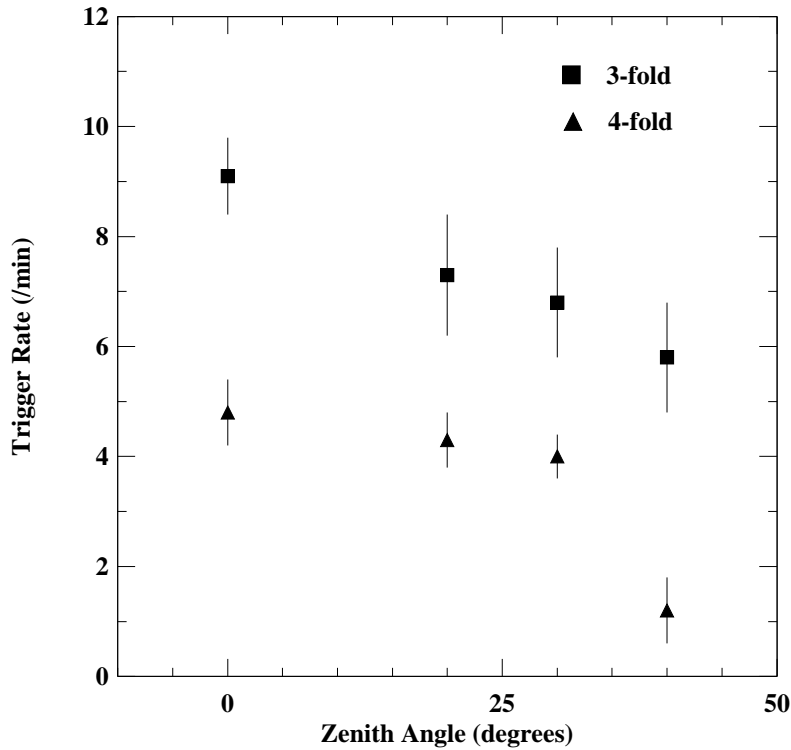,height=12.0cm}}
\caption{
Detector three and four-fold coincidence rates as a function
of air shower zenith angle.
The data were taken on November 4, 1994 at a discriminator threshold
of 125\,mV.} \label{fig:zenith}
\end{figure}

Fig.~\ref{fig:zenith} shows the variation in the trigger rate as a function 
of the air shower
zenith angle.
We saw a significant dispersion in the trigger rates at zenith
angles greater than $20^\circ$; the values shown in 
Fig.~\ref{fig:zenith} are
typical ones.
The trigger rate decreased monotonically with increasing zenith angle,
which is consistent with the expected increase in energy threshold
due to increased atmospheric depth.

\subsubsection{Pulse-Height Measurements}

\begin{figure}
\centerline{\ \psfig{figure=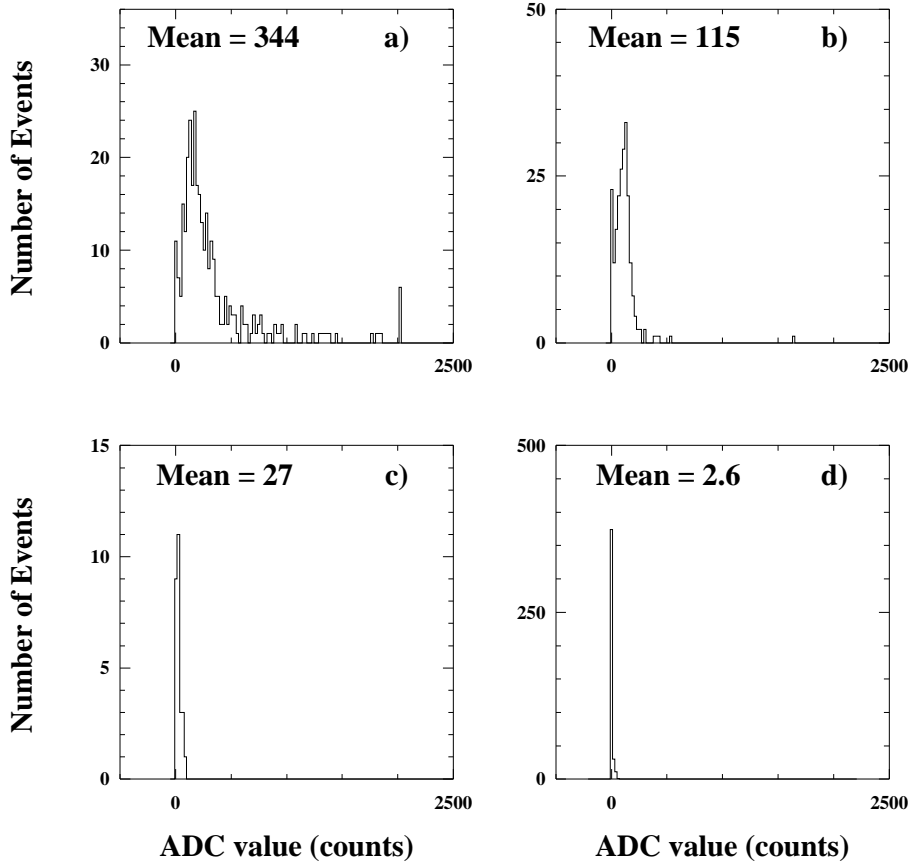,height=14.0cm}}
\caption{
Pulse height distributions for one of the four PMTs (PMT 2).
Histograms of the ADC count values are shown for:
a) four-fold coincidence events,
b) three-fold coincidence events where the PMT participated in the trigger,
c) three-fold coincidence events where the PMT did not participate
in the trigger,
and 
d) events in which the ADC was gated by a random trigger pulse.
The three-fold coincidence data shown in b) does not
include those events which were also four-fold coincident.
The data come from a ninety minute period on November 5, 1994, 
with the detector triggering on vertical air showers at
a discriminator threshold of 125\,mV.} \label{fig:pulses}
\end{figure}

The raw pulse-height distributions measured for one of
the four PMTs are shown in Fig.~\ref{fig:pulses}.
These data come from a ninety minute period on Night 2 at a discriminator
threshold of 125\,mV.
The mean PMT pulse-height in four-fold coincidence events was larger
than that in three-fold events, as expected.

Fig.~\ref{fig:pulses} also shows the pulse-height distributions for the
same PMT when it did not trigger the discriminator
in a three-fold coincidence event (Fig.~\ref{fig:pulses}c),
and when random gate pulses were applied to the ADC 
(Fig.~\ref{fig:pulses}d).
The mean pulse height for signals below the discriminator 
threshold is much less
than the mean for signals above threshold.
The mean in the former case is not zero, however, because in such events
the PMT detected Cherenkov signals, but
at levels below threshold.
The mean pulse-height for the ADC gated
randomly is close to zero, which is consistent
with the expected value of the ADC pedestal.

There is significant overlap between the distributions of 7b and 7c,
which shows that some pulses which participate in the trigger contain
less charge than pulses which are below trigger threshold.
We attribute this effect to the different pulse shapes produced in the
showers.
Some are of low (voltage) amplitude but are longer than pulses with equal
charge but higher amplitude. 
This lack of a constant pulse shape means that a voltage threshold is not
seen as a sharp edge in charge plots such as those shown in 
Fig.~\ref{fig:pulses}.

\begin{figure}
\centerline{\ \psfig{figure=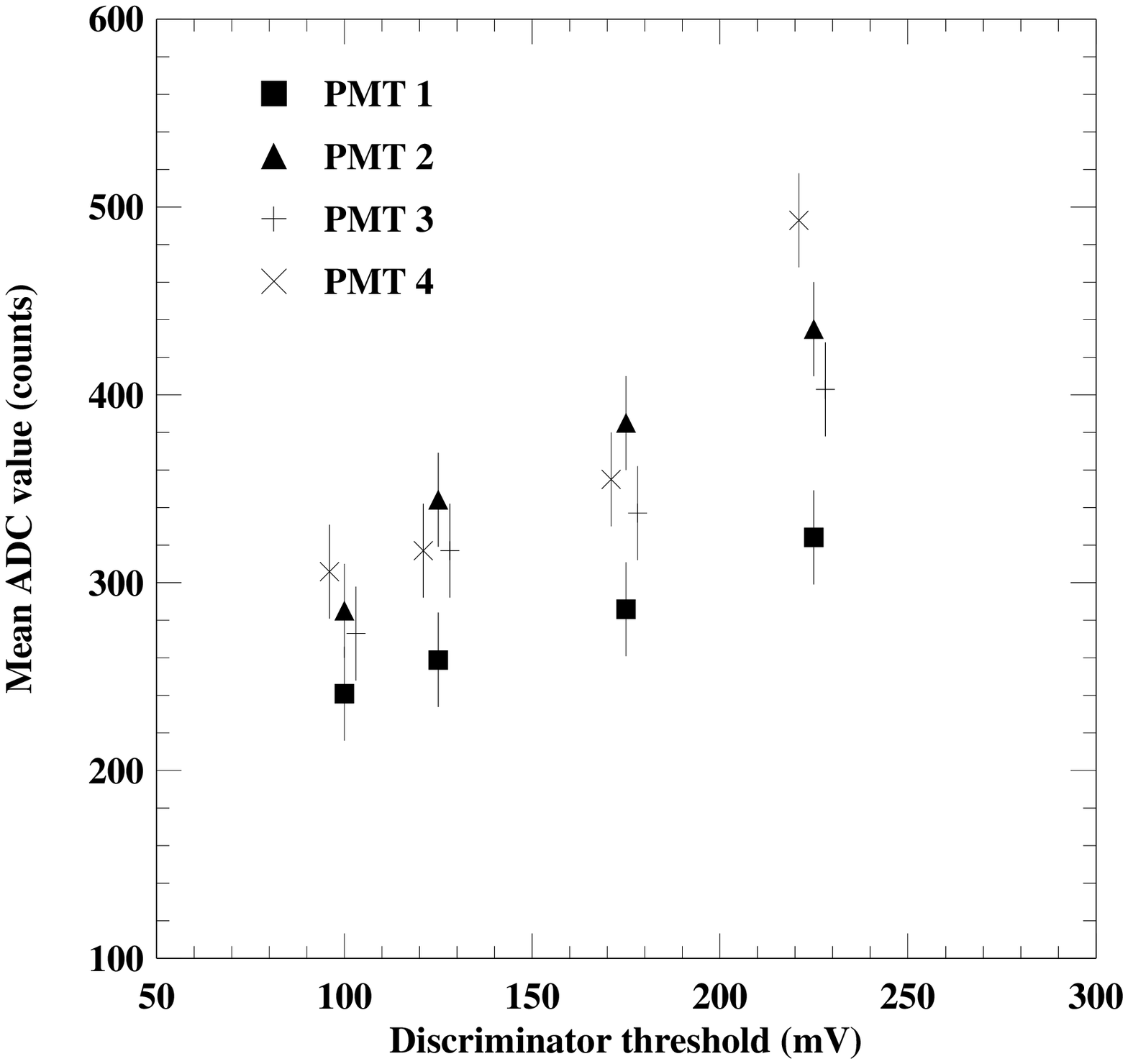,height=12.0cm}}
\caption{
Mean pulse-heights for each PMT as a function of discriminator
threshold for four-fold coincidence events.
The data come from a ninety minute period on November 5, 1994,
with the detector triggering 
on vertical air showers.} \label{fig:means}
\end{figure}

The mean pulse-height as a function of discriminator threshold
is shown in Fig.~\ref{fig:means} for each PMT in the detector.
Although there is some variation in gain from one PMT to another,
the average pulse-heights monotonically increase with increasing
threshold, as expected.

\subsubsection{Energy Threshold}
In the previous section, we have demonstrated that
the coincidence rates and pulse-height distributions 
are in agreement with expectations.
The data clearly show
that the detector was triggering on cosmic ray air showers.
Here we verify that the 
pulse-height data are consistent with what we expect
from cosmic ray showers triggering the detector at the
observed coincidence rate.
To verify the consistency of the data, 
we calculate the median energy for
cosmic rays triggering the detector
by two different techniques.
First, we compare the measured coincidence rates to the known 
integral cosmic ray flux, and second,
we use a simulation to reproduce the median 
observed pulse-height.
For these calculations, we use vertical air shower
data taken over a period of sixty minutes
on Night 1,
at a discriminator threshold of 125\,mV.

The mean three-fold coincidence rate for these data
was 9.1/minute; the mean four-fold rate was
4.8/minute.
In order to estimate the median cosmic ray energy of the detector,
we need to determine the expected coincidence rate as a function
of cosmic ray energy.
We estimate the integral cosmic ray flux, $\Phi\,( >E)$,
the effective collection area, $A$,  and the 
field-of-view of the detector, $\Omega$.
The expected coincidence rate, $R$, is then:

\begin{equation}
\ R\ =\ \Phi\,(>E) \cdot A \cdot \Omega \ . \label{eq:rate}
\end{equation}

\noindent
Our estimate of the integral cosmic ray flux is based on 
direct measurements 
made by balloon and satellite experiments \cite{ref:Swordy}.
We convert the differential spectra of the direct measurements
into an integral flux by estimating the power law
dependence of the spectra at 1 TeV.
We correct for the
fact that the direct techniques measure cosmic ray energies in units of
energy/nucleon whereas air shower detectors trigger on the
total energy.
Table~\ref{tab:nuclei} shows our estimates for the integral fluxes above 1 TeV
for the five major species of cosmic ray nuclei.

\begin{table}
\caption{
Integral flux estimates for various cosmic ray nuclei.
The flux above 1 TeV is expressed in the form
$\Phi$ (E $>$ 1 TeV)$ = \Phi_0\cdot {\rm E}^{-\alpha}$, where 
$E$ is the energy in TeV.
The fourth column lists the fractional percentage of each species
at 1 TeV.}
\label{tab:nuclei}
\vspace{10pt}
\begin{center}
\begin{tabular}{c|c|c|c} 
\hline
\ \ Species\ \ &\ \ $\Phi_0$ 
(/cm$^2$/sr/sec)\ \ & $\alpha$ & \ Fraction (\%)\ \\
\hline
H   &\ \ $5.87\times 10^{-6}$\ \ &\ \ 1.72\ \ & 35 \\
He  &\ \ $4.14\times 10^{-6}$\ \ &\ \ 1.68\ \ & 25 \\
CNO &\ \ $2.67\times 10^{-6}$\ \ &\ \ 1.62\ \ & 16 \\
NeS &\ \ $1.91\times 10^{-6}$\ \ &\ \ 1.61\ \ & 11 \\
Fe  &\ \ $2.13\times 10^{-6}$\ \ &\ \ 1.60\ \ & 13 \\
\hline 
\end{tabular}
\end{center}
\vspace{15pt}
\end{table}

From Table~\ref{tab:nuclei},
the total integral cosmic ray flux above 1 TeV is
estimated to be $(1.67 \pm 0.22) \times 10^{-5}$ particles/cm$^2$/sec/sr.
This flux represents the number of particles hitting the upper atmosphere
of the Earth per unit time per unit solid angle.
Our experiment, however, triggers on the air shower produced from
the cosmic ray interaction.
We must therefore account for the variation
in shower size for different cosmic ray nuclei.
Heavy nuclei interact sooner in the atmosphere and 
deposit a larger fraction of their energy into the hadronic cascade than 
do light nuclei.
These effects lead to a significant 
decrease in the Cherenkov light yield for air showers produced
by nuclei of increasing mass.

We estimate the amount of Cherenkov light expected for different cosmic
ray nuclei by means of the Monte Carlo simulation MOCCA 
\cite{ref:Hillas}.
For the hadronic and electromagnetic cascades that develop
from the primary interaction,
the trajectories of all particles with energies 
above 200 keV are followed
until they reach the ground or are absorbed.
The Cherenkov photons produced in the shower
are propagated to the ground, taking into account the 
wavelength-dependent absorption of the atmosphere from Mie and Rayleigh
scattering.
We determine the mean Cherenkov photon density at ground level
from those photons that 
are within a radius of $100\,$m of the shower axis and that arrive
within $40\,$nsec of the shower front.
The same average density in showers initiated by 
1 TeV protons is found in helium showers at approximately
1.5 TeV, nitrogen showers at 2.4 TeV, sulfur showers at 3.0 TeV,
and iron showers at 3.3 TeV.
We derive a corrected estimate for the integral
cosmic ray flux above 1 TeV of $(9.1 \pm 1.4) \times 10^{-6}$
particles/cm$^2$/sec/sr, where the flux has been corrected
for composition effects (i.e. it describes the total flux of
particles that would yield showers having
the same Cherenkov photon density as 
that in a shower produced by a 1 TeV proton).

We estimate the field-of-view for the detector to be
$(7.9\pm 0.5) \times 10^{-5}\,$sr, and 
the collection area to be
$(3.1 \pm 0.8) \times 10^{8}\,$cm$^2$.
The field-of-view is determined from a calculation which
convolutes the measured heliostat
beam size with the dimensions of the camera box aperture.
The collection area is determined from the simulation.
The error in the collection area results largely from
our uncertainty in the value of the energy threshold and the fact
that the collection area depends on threshold.
We use 
Eq.~(\ref{eq:rate}) to
determine a median energy for
the detector for proton showers of
$1.24 \pm 0.19\,$TeV for the three-fold coincidence data,
and $1.79 \pm 0.25\,$TeV for the four-fold coincidence data.
This determination assumes
an overall cosmic ray integral spectral index of
1.66.

We also determine the median detection energy from the
pulse-height distributions derived from the same data set.
Assuming that the 
observed Cherenkov photon density on the ground is proportional to 
the primary particle energy,
the detected number of photoelectrons for each tube, 
$N_{\rm{pe}}$, can
be represented as:

\begin{equation}
\ N_{\rm{pe}}\ \ =\ \ 
y_{\rm{p}}\,E\,\epsilon\,A\ , \label{eq:ph}
\end{equation}

\noindent where $y_{\rm{p}}$ is the expected Cherenkov
photon density on the ground per unit of proton energy,
$E$ is the energy, 
$\epsilon$ is the
efficiency of photon detection, and $A$ is the projected
area of each heliostat.
Using the simulation, we estimate that 
$y_{\rm{p}} \sim 56\,$photons/m$^2$/TeV
near 1 TeV
and that $\epsilon = (7.3\pm 1.5)\times 10^{-3}$.
The value for $\epsilon$ includes 
the wavelength-dependent reflectivity of the heliostat mirror (0.51),
the fraction of the heliostat image within the aperture of the Fresnel
lens (0.14), the transmission of the Fresnel lens (0.85), the fraction of
the Fresnel image collected by the PMT (0.80), and the quantum efficiency of
the PMT (0.15).
The projected area of the heliostat is 32.7$\,$m$^2$ for these data.

The scale conversion between the number of photoelectrons
incident at the PMT face and the measured pulse-height (in ADC
counts) was determined by measuring the pulse-height distribution of 
a PMT from the camera
viewing a 0.6\,m x 0.6\,m x 0.64\,cm piece of acrylic scintillator.
The photoelectron yield of the scintillator was measured as part
of a different experiment 
\cite{ref:Borione}.
This procedure yielded a conversion factor of $9.1\pm 1.1$
ADC counts/photoelectron.
Using this factor,
the median pulse-height 
was $14.9\pm 1.8$ photoelectrons for the
three-fold coincidence data, and
$21.3\pm 2.6$ photoelectrons for the four-fold coincidence data.
Using Eq.~(\ref{eq:ph}), we can therefore estimate a median energy
of the detector for proton showers of
$1.12 \pm 0.23\,$TeV for the three-fold coincidence data,
and $1.60 \pm 0.32\,$TeV for the four-fold coincidence data.
The errors in the energy estimations are dominated by our
uncertainty in the value of $\epsilon$.
Table~\ref{tab:energies} 
shows the median energy estimations by the two different methods.
Within the uncertainties, the
two methods agree.

\begin{table}
\caption{
Median proton energies for the three and four-fold coincidence
data by the two different methods described in the text.}
\label{tab:energies}
\vspace{10pt}
\begin{center}
\begin{tabular}{c|c|c} 
\hline
\ \ Method\ \ &\ \ E (3-fold)\ \  &\ \ E (4-fold)\ \ \cr
\hline
\ \ Rate\ \ &\ \ $1.24 \pm 0.19$ TeV \ \ &\ \ $1.79 \pm 0.25$ TeV \ \ \cr
\ \ Pulse-height\ \ 
            &\ \ $1.12 \pm 0.23$ TeV \ \ &\ \ $1.60 \pm 0.32$ TeV \ \ \cr
\hline 
\end{tabular}
\end{center}
\vspace{15pt}
\end{table}

The lowest energy threshold obtained for the detector was
on the first night of operation at a discriminator
threshold of 90 mV.
In this configuration, the three-fold coincidence rate was
14.6/minute.
From these data, we estimate a median detection energy for
proton showers of $940 \pm 130$ GeV.
Since 500 GeV gamma-rays produce showers with a 
comparable amount of Cherenkov radiation 
to showers initiated by 1 TeV protons,
the median energy for the detector described in this
paper was near or below 500 GeV for gamma-ray showers.

In addition, during the tests, we made no attempt to explore how low
in energy threshold the detector would reliably operate.
Since signal-to-noise was not a problem in our tests, the detector could
clearly have operated at lower energies.
Such operation was not attempted experimentally due to lack of time.
We can estimate the minimum energy threshold from the
observed PMT single counting rates and from laboratory measurements 
using the same type of PMT \cite{ref:PMT_Threshold}.
Assuming a maximum  accidental trigger rate of 10~Hz, we estimate
a minimum single PMT threshold of approximately six photoelectrons.
This PMT threshold translates into an overall 
energy threshold of $\sim 450\,$GeV for vertically incident proton showers
using  this initial prototype detector.

\section{Conclusions}

We have built a prototype atmospheric Cherenkov detector using
solar heliostat mirrors as the primary reflecting element.
The detector was operated successfully in November, 1994.
The observed detection rate of cosmic ray air showers was consistent
with our expectations, 
and with a median proton energy of 1 TeV.
We are now developing a first-generation experiment that will 
employ
up to fifty heliostats and several large area secondary collectors.
This experiment could take initial data in 1997.



\vspace{5mm}
\noindent {\bf Acknowledgements}
\vspace{3mm}

We would like to acknowledge the cooperation and help of
Southern California Edison in the tests at the Solar Two Power Plant.
We thank Charles Lopez, Roy Takekawa, and Robert Edgar, and
gratefully acknowledge the help of 
Mark Chantell and Anthony Miceli.
This work was supported by the National Science Foundation,
the Institute of Particle Physics of Canada, and
the California Space Institute.
RAO wishes to acknowledge the support of the W.W. Grainger Foundation
and the Alfred P. Sloan Foundation.



\begin{thebibliography}{99}
\bibitem[14]{ref:Heliostats}
The identification numbers 
of the heliostats used were
1401, 1403, 1405, and 1407.
Their 
locations were (10.6,189.6) m,
(23.7,188.4) m, (36.7,186.3) m, and (57.1,181.1) m,
respectively,
in a coordinate system with (x) pointing east and 
(y) pointing north.
\bibitem[19]{ref:PMT_Threshold}
In the laboratory, we used a constant intensity light source to
illuminate a PMT.
We adjusted the source intensity so that the single PMT counting
rate matched the observed detector rate at the
same PMT discriminator threshold used in the detector.
We then measured the dependence of the PMT counting rate versus
discriminator threshold.
\end{thebibliography}
\end{document}